\documentstyle[aps,multicol,epsfig]{revtex}
\author{Mendeli H. Vainstein and Jeferson J. Arenzon\footnote{Corresponding
author: {\tt arenzon@if.ufrgs.br}}} 
\address{Instituto de F\'\i sica \\
Universidade Federal do Rio Grande do Sul \\ 
CP 15051 \\
91501-970 Porto Alegre RS -- Brazil}
\title{Disordered Environments in Spatial Games}
\date{\today}

\newcommand{\bea}{\begin{eqnarray*}}
\newcommand{\eea}{\end{eqnarray*}}

\def\fiwi{5.5cm}

\begin{document}
\maketitle
\begin{abstract}
The Prisoner's dilemma is the main game theoretical framework in
which the onset and
maintainance of cooperation in biological populations is studied.
In the spatial version of the model, we study the robustness of 
cooperation in heterogeneous ecosystems in
spatial evolutionary games by considering site diluted lattices.
The main result is that due to disorder, the fraction of cooperators in
the population is enhanced. Moreover, the system presents a 
dynamical transition at $\rho^*$, separating
a region with spatial chaos from one with localized, stable
groups of cooperators.
\end{abstract}

\begin{multicols}{2}   
\narrowtext

\section{Introduction}

Game theoretical methods have been
applied, in a quite sucessfull way, to several different
fields~\cite{Axelrod84,LuRa85}, biology being one of the most
successful branches~\cite{Smith82,Weibull95,Hamilton64,DuRe98}. 
In these evolutionary games, 
rewards are translated in 
terms of subsequent reproductive success, a more natural 
and clear scale than the rationality dependent ones used in, for 
example, economy. Of particular interest is the 
emergence and sustainability of cooperation, what has
attracted a lot of attention as it 
poses a difficult problem from an evolutionary point of view
since cooperators, by their nature, have their
fitness decreased while interacting with defectors. Theories advanced
to explain the evolution of cooperation usually consider either
kin selection~\cite{Hamilton64}, reciprocal altruism~\cite{Trivers71,AxHa81} 
or group 
selection~\cite{Williams66,Wilson80,Donato96,DoPeSe97,SiFo99}, 
the separation between them not always being clear.
Group foraging and young raising~\cite{HePa95}, alarm 
calls~\cite{Smith65,BeLa01}, 
bacteria-infecting viruses~\cite{TuCh99}, predator inspection
by fish~\cite{Milinski87}, birds female-female 
cooperation~\cite{HeTr98}, cleaner fishes~\cite{Grutter99},
are but a few examples of such acts in biological populations,
ranging from simple to complex organisms. A number of other
examples may be found in ref.~\cite{DuRe98}.

The Prisoner's Dillema game is 
generally studied as an archetypical model for reciprocal altruism. 
Each of the two players either cooperates ($C$) or
defects ($D$), without knowledge of the opponent's strategy.
The result depends on the mutual choice and is given by
the payoff matrix whose elements are:  a reward $R$ (punishment $P$)
if both cooperate (defect), $S$ (sucker's payoff) and $T$ (temptation)
if one cooperates and the other defects, respectively.
Moreover, these
quantities should satisfy the inequalities $T>R>P>S$ and
$2R>T+S$. It is clear that, independent of the opponent's
choice, defecting is always the best bet. Thus, two basic
evolutionary problems are $i$) the onset of cooperation, that is, under
which conditions a given cooperative behavior can invade a
population of defectors and, once established, $ii$) its stability, that is,
under which conditions the population of cooperators is uninvadable.
In other words, we are looking for Evolutionary Stable Strategies
(ESS)~\cite{Smith82,Weibull95,HoSi98} where there is stable mutual 
cooperation between individuals. 
In a random mating, infinite population of asexual (haploid) elements,
where two pure strategies are present (cooperators $C$ and
defectors $D$), it can be shown that defecting will be
the most rewarding strategy (an ESS). If instead of only interacting
once or a known finite number of times, the players have a probability
$w>0$ of meeting again in the
next round and remember the chosen strategies in previous encounters,
more complex rules may be devised~\cite{Axelrod84}. For instance,
in a round robin tournament, Axelrod~\cite{Axelrod84} found that
tit-for-tat (TFT) (starts cooperating and, after that, do what the opponent did
in the previous step) was the most successful strategy, but other
successful strategies have also been found~\cite{NoSi93}. Nevertheless,
TFT cannot invade a population of defectors before reaching a
minimum population, what can be achieved in several 
ways~\cite{AxHa81,FeMi96,DoBlAc97}.

Following a suggestion of Axelrod~\cite{Axelrod84},
Nowak and May~\cite{NoMa92,NoMa93,NoBoMa94b} have shown how,
cooperation can arise
even with only pure, memoryless strategies, like always cooperate
or always defect in the presence of spatial structure.
 They considered a deterministic cellular automaton 
consisting of a square lattice with
near and next nearest neighbors interactions and self interaction.
Since the agents are spatially 
localized, they are more likely to interact only with their nearest
neighbors, differently from the standard, mean field-like approach, 
that consider an infinite, random mixing population.
 The reasons for this are manyfold: individuals usually
occupy well-defined territorial regions, individuals do not move far from 
their places of birth (population 
viscosity~\cite{Hamilton64}), 
interactions occur in places where animals
usually meet such as water ponds, etc. After the combats,
each player compares its total payoff with the ones of its
neighbors and changes strategy, following the one with the
greatest payoff among them. For a certain range of values of the
payoff matrix, very complex spatial patterns show up with cooperators and
defectors coexisting (spatial chaos). 
In these structured populations, cooperative strategies can build clusters
in which the benefits of mutual cooperation can outweight losses against
defectors, maintaining the population of cooperators stable. 
Irrespective of the initial state that
may be chosen with either only one initial defector or a fraction
of randomly distributed defectors, the asymptotic density depends
only on the payoff matrix parameters. 
The actual values depend on the
neighborhood chosen for the dynamics and whether self-interactions
are included or not~\cite{NoMa92,NoMa93}.
This cooperation enhancement effect due to the
spatial structure is also seen in other games.
For instance, in the spatial version of the Hawk-Dove (HD)
game~\cite{NoMa93,KiDo96,KiDo98}, the density
of doves (analogous to cooperative individuals) is increased by the
spatial structure, although in this game there would exist polymorphism
even in the absence of spatial effects.
In a similar way, several strategies are more successful in a spatially 
structured population
than otherwise~\cite{LiNo94,BrKiDo99}.


In real populations, however, not all individuals 
interact the same number of times
either due to the non sincronous character of the interaction 
or to the environment structure
that prevents some of the contacts. Both the spatial and
temporal aspects of the environment affect the interactions
between individuals, being central issues in ecological and
evolutionary theory. Thus, a natural question
arises: how is cooperation affected by this inherent inhomogeneity?
 In other words, how robust is cooperation in the
presence of disorder? 
In this paper we try a simple approach to this
question by considering a regular lattice where
some sites are empty. Dilution  can be either quenched (fixed) or
annealed (evolving). In the latter case, individuals may diffuse in
the lattice, what will be considered in detail in a future 
work. Quenched vacancies (or defects) 
may account for the presence of environmental features ({\em e.g.}, 
geographical) in the game, making some 
individuals have less neighbors than others. 
Deviations from the ordered lattice can also
be achieved in several other ways. For instance, by allowing
that some of the individuals also interact with distant ones, 
in a small-world network fashion~\cite{WaSt98,AbKu01}.
In~\cite{NoBoMa94b,NoBoMa94a}, a diluted lattice was used with
individuals interacting inside a region of radius $r$, and
persistent polymorphism of $C$'s and $D$'s was found, unless
$r$ was made too big, reaching the long range connectivity 
associated to mean field behavior, where the defector population
dominates. Nevertheless, their treatment was quite brief
and several interesting dynamical behaviors have passed 
unnoticed, as well as the important issue of whether disorder
enhances or not the fraction of cooperators in the population.

Here we show that depending on the amount of disorder, cooperation
can be enhanced, there being a point where a dynamical transition
settles in, separating the region with spatial chaos from the one
with localized groups of cooperators.
The paper is organized as follows. In section~\ref{section.model},
the diluted version of the spatial prisoner's game is presented and
the order parameters are defined. In section~\ref{section.results}
the main results are presented and, finally, in 
section~\ref{section.conclusions},
we present the conclusions and comments.

\section{The Model}
\label{section.model}

We consider the spatial version of the prisoner's
dilemma~\cite{Axelrod84,NoMa92,NoMa93,NoBoMa94b,LiNo94,NoBoMa94a,SzTo98,ChOl99,SzAnSzDr00}, 
placing the interacting elements
in the vertices of a $d$-dimensional array, usually a hypercube,
with periodic boundary conditions. The results presented here are
for $d=2$.
Differently from the original case, we allow 
 that some of the sites may be empty. To describe
the occupation of the $i$-site ($1<i<N=L^d$,  where
$L$ is the system linear dimension) we take $n_i$ to be either 1, if 
the site is occupied or 0, otherwise. 
In every generation, each individual assumes an
unchangeable strategy from a set $\Omega$ and, following ref.~\cite{NoMa92},
we only consider the simplest case of pure strategies, $C$ and
$D$, represented by the variable $S_i=\pm 1$, 
respectively. In each step (generation), the $i$-th individual
($n_i\neq 0$) combats
with all other elements inside a given neighborhood ${\cal V}_i$, 
and accumulates
a payoff $p_i$, depending on the chosen strategies, according to the 
reduced payoff table for the prisoner's game~\cite{NoMa92}: 
$R=1$, $P=S=0$ and $T=b>1$, reducing the problem to only one parameter
(besides the density). For the case considered here, where the neighborhood
${\cal V}_i$ is restricted to the nearest neighbours of site
$i$ and no self interactions are included, both $C$ and $D$ coexist
in the region $4/3<b<3/2$ in which also the number of active sites (see
below) is
large. For $1<b<4/3$ cooperators dominate while for $b>3/2$, defectors
are dominant.
The player's payoff is a measure of its reproductive success: when 
reproducing, the $i$-th
element compares its own payoff with all $j\in {\cal V}_i$
and changes to the strategy of the site that has the greatest
payoff in $\{i\}\cup {\cal V}_i$.
In this way, the global density is kept fixed since no empty site will
be ever filled. 

To characterize the macroscopic behavior of the system we introduce
two order parameters.
Let $\rho_c(t)$ represent the fraction of cooperators at a given
time:
\begin{equation}
\rho_c(t) = \frac{1}{2N} \sum_{i=1}^N (S_i+1)n_i 
\end{equation}
where $N=L^d$ is the total number of sites. Clearly, $\rho_d(t)$,
the defectors density, is $\rho_d(t)=\rho-\rho_c(t)$,
where $\rho$ is the total density. 
Since we are 
interested in the long time regime, and the results
depend on the choice of the frozen empty sites, 
we define the order parameter as the average over time 
($\langle \ldots\rangle$) and over 
the realizations of the disorder ($\overline{\cdots}$) of
the asymptotic cooperators density,
$\rho_c = \langle \overline{\rho_c(\infty)} \rangle$, 
for large $N$. Sometimes it is
more useful to have the relative cooperators density, $\rho_c/\rho$.
Thus, $\rho_c=0$ means that the population was fully invaded by defectors
and $\rho_c=\rho$, by cooperators. An intermediate case, 
$0<\rho_c<\rho$
in which both strategies coexist is also possible. Moreover,
it is interesting to know the fraction of active sites, that is, the fraction
of elements that change strategy in time:
\begin{equation}
\rho_a(t) = \frac{1}{2N} \sum_{i=1}^N (1-S_i^t S_i^{t-1})n_i 
\end{equation}
This defines our second order parameter, 
$\rho_a = \langle \overline{\rho_a(\infty)} \rangle$.
Thus, $\rho_a=0$ means that all sites are frozen and $\rho_a=\rho$ 
that all elements are changing strategy.

\section{Results}
\label{section.results}

\begin{figure}
\epsfig{file=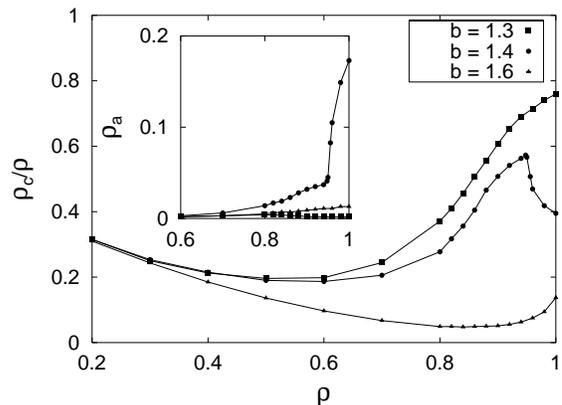,width=\fiwi,angle=270}
\caption{Asymptotic fraction of cooperators, $\rho_c/\rho$, as a
function of the lattice occupation, $\rho$ for
several values of $b$. The initial state has $\rho/2$ 
cooperators, the lattice size is $L=100$ and 
averages are taken over 100 samples.}
\label{nodiff}
\end{figure}

In fig.~\ref{nodiff}, the asymptotic cooperator density 
$\rho_c/\rho$ is
plotted against the total occupation of the lattice,
$\rho$.
If the occupation fraction is near zero, almost
all sites are isolated and do not change strategy since there
is no combat at all, the asymptotic
density being the same as the initial one: $\rho_c=\rho_c(0)=\rho/2$.
Indeed, for small $\rho$, the curves for different values of 
$b$ merge.
As the density increases, the probability of occurring pairs 
of occupied sites increases and, irrespective of the value of
$b$, all $CD$ pairs will become $DD$, what can be seen as
a decreasing curve of $\rho_c$ from the origin. Around 
$\rho\simeq 0.3$, as
the number of interacting individuals increases due to the
increasing clusters sizes, the dynamics will define the fate
of each cluster and the curves for different
values of $b$ depart: the higher $b$ is, the better is for
defectors and the lower is the corresponding cooperator curve. 
Still further, when clusters
of cooperators have a reasonable probability to be formed, their
density start to increase recovering, as the total 
density approaches unity, the Novak-May results~\cite{NoMa92}. 
The most interesting case happens if $b$ is in the active region,
$4/3<b< 3/2$, where the behavior is not monotonic and a sharp decrease
in $\rho_c$
appears near $\rho^*\simeq
0.95$. At the same point, the fraction of active
sites presents a sharp increase, as can be seen in the inset
of fig.~\ref{nodiff}. Notice that for the others regions, the
behavior is almost uniform, with a very small number of active
sites.
Interestingly enough, depending on the region of $b$,
the optimum density that
maximizes the fraction of cooperators occurs at different values: 
for $b<4/3$, the more individuals, the
better, and their maximum occurs at $\rho=1$; above
$b=3/2$, on the other hand, the less occupied the network, the
better, due to the exploitation
by defectors; the maximum occurs in the limit $\rho
\to 0$. In these cases, respectively, cooperators and defectors
are in advantage when interacting.
In the intermediate, active region, the behavior is non trivial
and the maximum happens
at the transition point $\rho^*$. 
Remarkably, this point is much above the site percolation
transition, that for a square lattice is located at $\rho_{perc}\simeq
0.59$~\cite{StAh94}: although there is a connected, infinite cluster,
regions of active sites are bounded (pinned) to small regions due to the
presence of defects, as will be shown below.
Below the transition, the approach to equilibrium is exponentially
fast, $\rho_c(\infty)-\rho_c(t)\sim \exp(-t/\tau)$, $\tau$
diverging as one approaches the point $\rho^*$. 
 This is because most of the interactions occur 
inside of the localized groups.
Also, if the lattice is completely filled,
the approach is fast. On the other hand, just above the critical
value, the presence of defects, and the fact that only some of
the groups are depinned makes the approach to equilibrium 
extremely slow, power-law like, what is reminiscent of disordered systems.

\begin{figure}
\epsfig{file=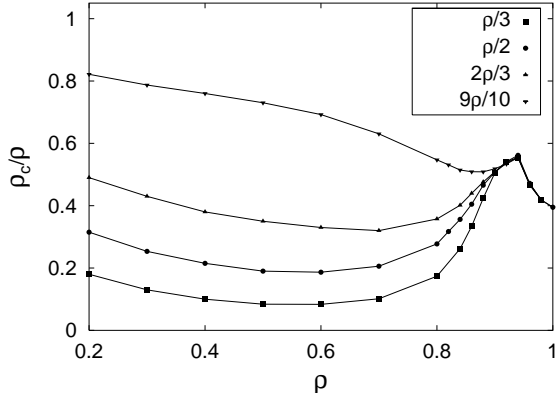,width=\fiwi,angle=270}
\caption{The asymptotic fraction of cooperators for several different
initial densities of cooperators, $L=100$ and $b=1.4$.}
\label{nodiff2}
\end{figure}

In the active region, for different initial concentrations of 
cooperators, it can
be seen that almost up to $\rho^*$, the asymptotic
density depends on the initial state, as shown in
fig.~\ref{nodiff2}. For small $\rho$, since the
clusters are independent, obviously the higher  
the initial fraction of cooperators is,
the higher $\rho_c$ will be.
 Notice that the optimum density for cooperation is
a function of the initial density of cooperators: for initial
concentrations below the height of the peak at $\rho^*$,
$\rho_c(0)\lesssim 0.54\rho$, $\rho^*$ is the ideal 
total concentration for cooperators, while above, the smaller
is $\rho$, the better.
Slightly below $\rho^*$ all curves merge because the
cluster boundaries are no longer completely pinned and the 
memory of the initial state is washed out.

\begin{figure}
\epsfig{file=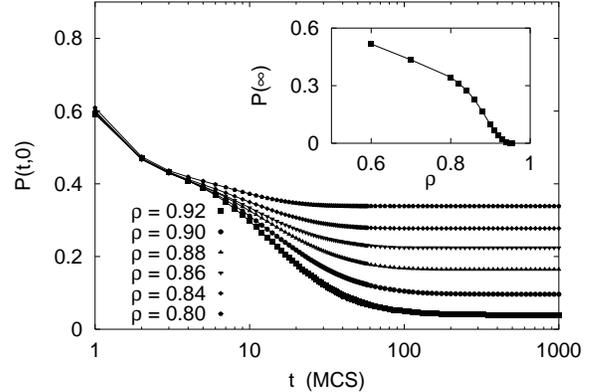,width=\fiwi,angle=270}
\caption{Persistence $P(t,0)$ for $L=100$, $b=1.4$  and several 
total densities $\rho<\rho^*$ as a function of time (measured in
Monte Carlo steps). Notice that these densities are 
far above the percolation threshold.
As $t\to\infty$, all curves attain the
$\rho$-dependent plateux $P(\infty,0)$, shown in the 
inset as a function of the lattice
occupation. The behavior of $P(\infty,0)$ near $\rho^*$ is
power law, with an exponent aproximately 2.2.}
\label{persistence}
\end{figure}

\begin{figure}
\epsfig{file=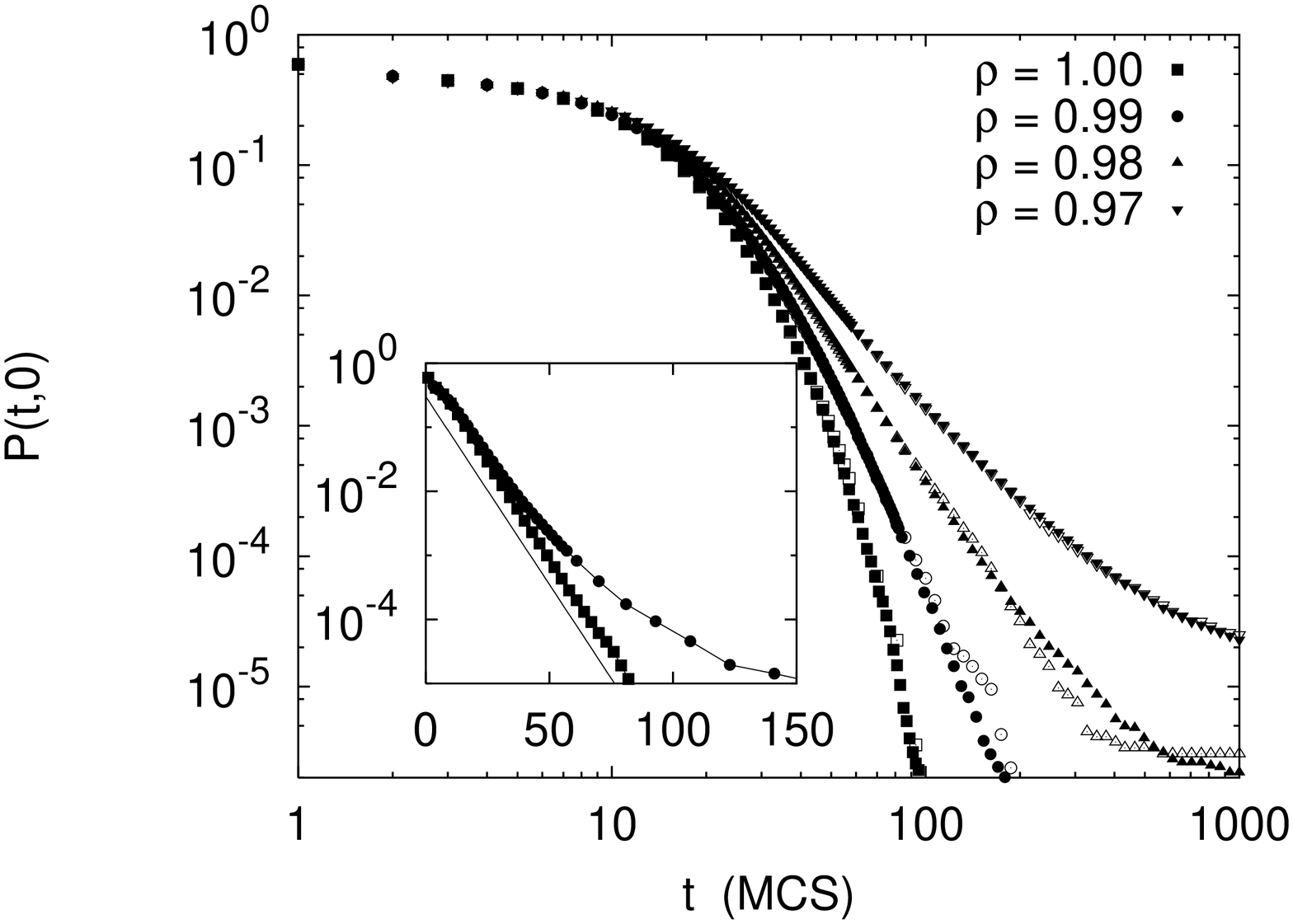,width=\fiwi,angle=270}
\caption{Persistence $P(t,0)$ for $L=200$ (empty sites) and
500 (filled ones), $b=1.4$  and several 
total densities $\rho>\rho^*$ as a function of time (measured in
Monte Carlo steps). Differently from the $\rho<\rho^*$
case, here the persistence goes to zero (taking into account
only sites in the infinite cluster). For $\rho=1$ it vanishes 
exponentially (see the semi-log plot in the inset), 
while near this point an initial
exponential decay can be seen. Below $\rho=1$, the longtime
behavior follows a power law, unless we are too close to
$\rho^*$, where a crossover behavior to the plateux is noticed.}
\label{persistence_maior}
\end{figure}

\begin{figure}
\begin{center}
\epsfig{file=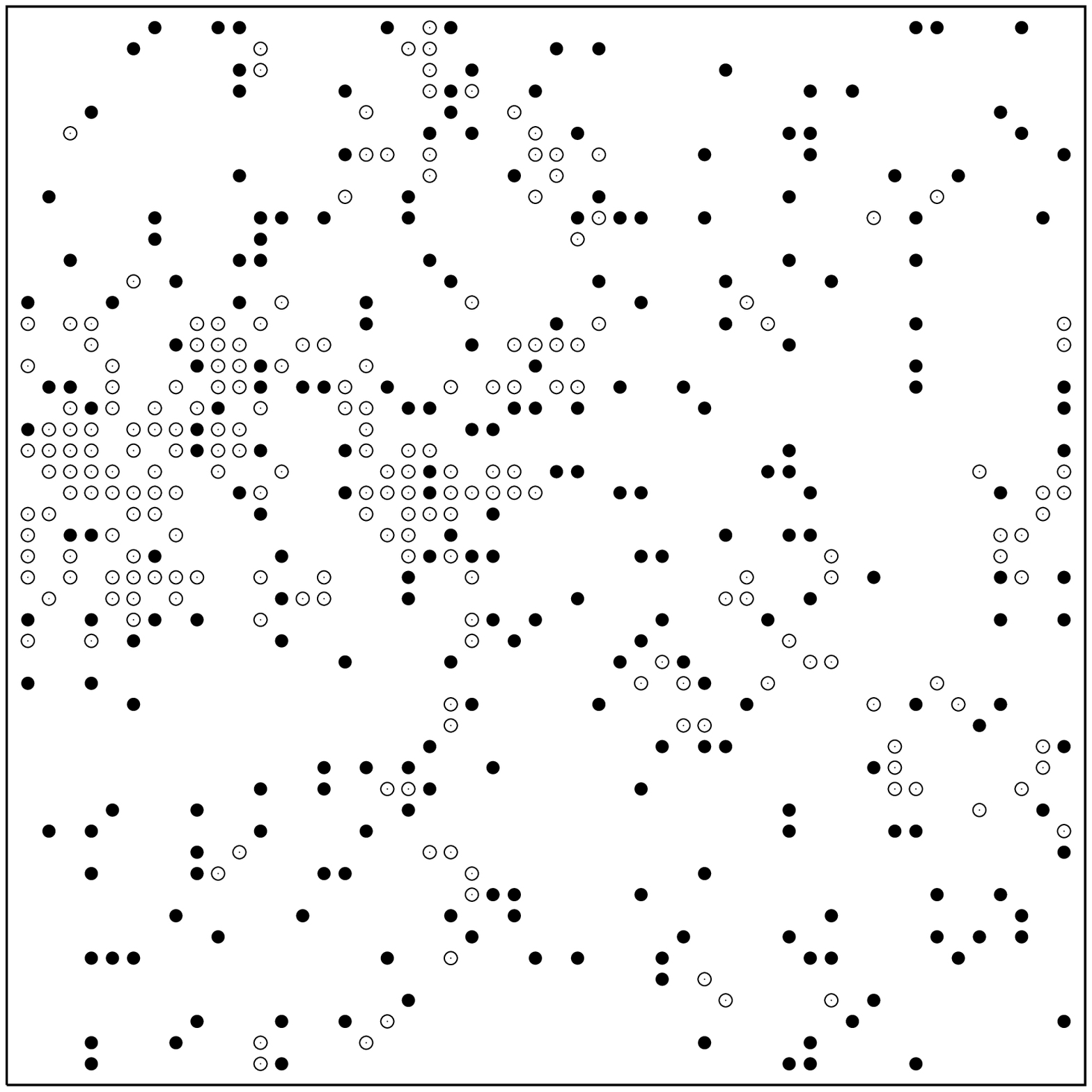,width=5.5cm,angle=270}

\epsfig{file=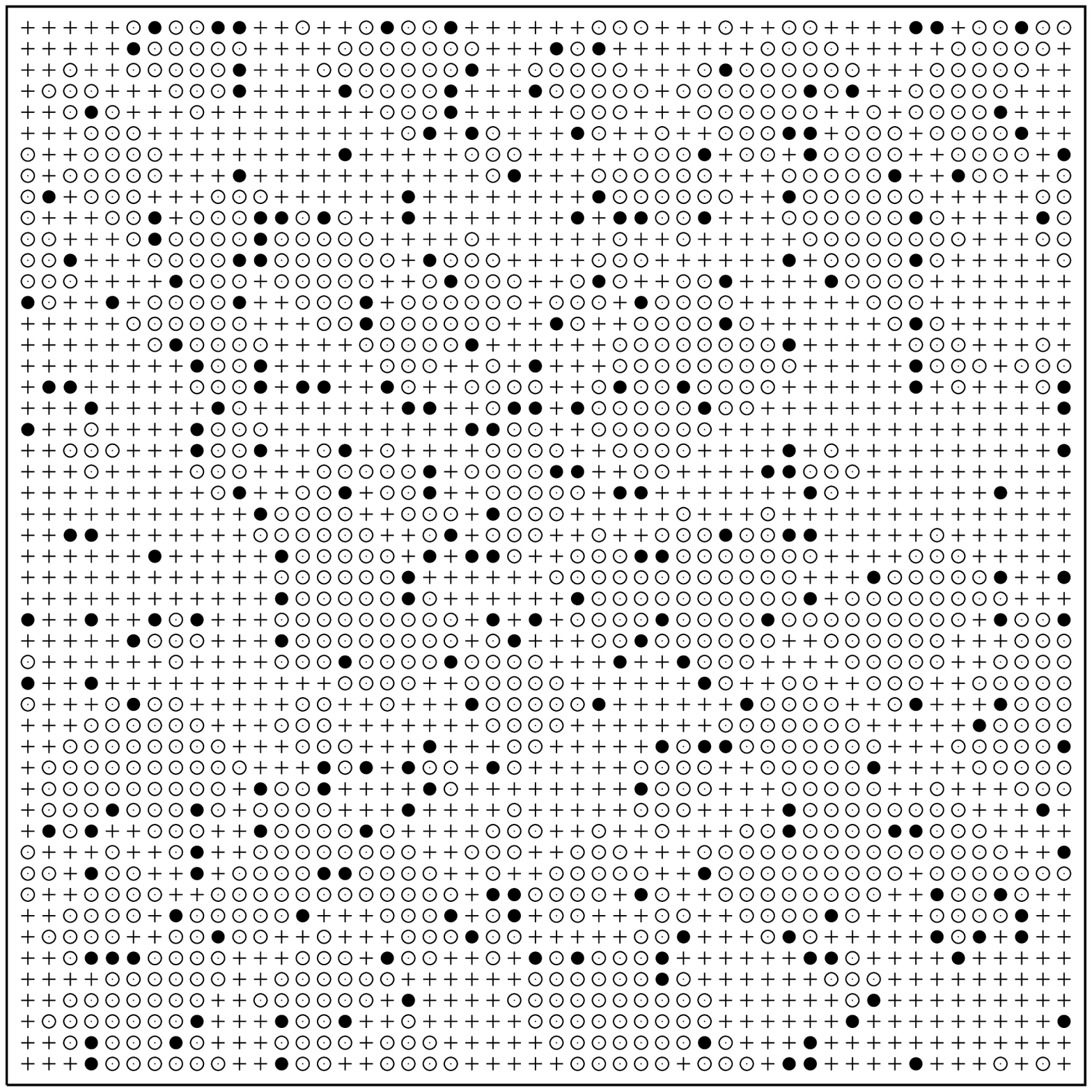,width=5.5cm,angle=270}
\end{center}
\vspace{0.5cm}
\caption{Snapshot showing the sites that have not changed strategy
after $10^5$ steps for $L=50$ and total density 0.9 (top).
The figure at the bottom shows, at the same time, the
cooperators (empty circles) and defectors (crosses).
The black points are the empty sites. Those sites that are blocked 
will remain blocked forever.}
\label{persistence_lattice}
\end{figure}

Although both cooperators and active sites densities are useful
parameters to describe the system asymptotic behavior, they are not 
sufficient in order to understand the complex dynamical behavior
that arises in the presence of disorder. To clarify what is going on at 
the transition point, we also measured the persistence, the fraction
$P(t,t_w)$ of sites that do not change strategy between an initial 
waiting time $t_w$ and the time 
$t>t_w$~\cite{DeBrGo94,BrDeGo94,Stauffer94,Majumdar99}, as can be seen 
in figs.~\ref{persistence} and \ref{persistence_maior}.
Differently from the fraction of active sites, the persistence
is a very complex measure that depends on the whole time history
of the system since $t_w$.
For example, if the persistence does not go to
zero, we know that there is a fraction of sites that flips only
finitely many times~\cite{DeOlSt96,NeSt99b}
 (blocking) and domain wall movements are constrained
(pinning). That is precisely what happens for $\rho<\rho^*$, as
shown in fig.~\ref{persistence}: after an initial decrease,
the persistence attains, for large times, a plateu whose value depends 
both on $\rho$ and $t_w$. Denoting  this plateu by $P(\infty,t_w)$, we 
notice that it goes to zero at $\rho^*$,
as shown in the inset of fig.~\ref{persistence} for $t_w=0$, signalling
a depinning transition. In the critical region, the behavior is 
power-law: $P(\infty,0)\sim  (\rho^*-\rho)^{2.2}$. It is important
to notice that the contribution from isolated sites to the plateau
is small, and most of the sites forming the plateau come from the
infinite cluster. 
At $\rho^*$ and above, the interfaces are
no longer constrained, the number of active sites suffer
a sudden increase and all sites eventually change strategy.
It has been shown for several models~\cite{NeSt99b,Jain99}
that in the presence of disorder (but also in some homogeneous
systems), the decay to the plateau, $P(t,0)-P(\infty,0)$,
is exponential~\cite{NeSt99b,Jain99} at large times.
The same behavior is found here for a non Hamiltonian model.
In fig.~\ref{persistence_lattice} we present a snapshot
of the system below $\rho<\rho^*$, showing both the empty
and the pinned sites, as well as their strategies. This
configuration is almost stable, many of the cooperator groups
shown no longer change, and the small amount of active sites
are either confined to a few regions or belong to one
of the blinkers present. This has to be compared with the
spatial chaos region above $\rho^*$, where there is no
blocking and every spin flips infinitely many times.
Since we are near the transition, the
quantity of pinned sites is quite reduced.

For densities $\rho>\rho^*$, considering only elements
that belong to the infinite cluster, the persistence goes to zero,
as is shown in fig.~\ref{persistence_maior}. For
$\rho=1$, the behavior is exponential, as shown in the
inset of fig.~\ref{persistence_maior}. Notice, also in the
inset (for $\rho=0.99$), that as we dilute the lattice, only the initial
behavior is exponential. After this initial exponential decrease, whose
length decreases as $\rho$ departs from 1, the behavior follows a
power law, $P(t,0)\sim t^{-\theta(\rho)}$, and as we approach
$\rho^*$ from above, the plateau starts developing. The exponent
$\theta(\rho)$ is the persistence exponent and has non trivial
values, depending on the total density. 
For example, $\theta(0.98)\simeq 3.6$
and  $\theta(0.99)\simeq 5.7$.

Below the critical value, the groups of cooperators are localized and
their borders cannot move because of the presence of the defects.
We say the groups are pinned by the environment. 
This explains why the fraction of cooperators
is larger in this regime. As the number of defects decreases, the
groups start to become depinned, the interfaces start moving and
interference effects between the groups appear, explaining the
sudden increase in the number of active sites. From the persistence
data, we can see that below the critical value, there is a constant
fraction of sites in the infinite cluster
that never change strategy, meaning that the
clusters are localized. As we approach the deppining transition,
some of the groups are depinned and the persistence decreases.
Since some of the groups are still pinned, we observe a constant
plateau that decreases as $\rho$ approaches $\rho^*$.
Notice that below the transition, the number of active sites
is not zero, but small, due to blinking sites.

\section{Conclusions}
\label{section.conclusions}

One of the main features of the Prisoner's dilemma, responsible
for its widespread use in the problem of cooperation,
is its robustness. Here we have shown that the system is able
to sustain cooperation even under the presence of lattice disorder,
what was already suggested in ref.~\cite{NoBoMa94b}. However,
what had been unnoticed was the dynamical transition as a
function of the amount of quenched disorder as well as
the fact that the disorder may enhance cooperation. Dilution changes 
the scenario presented by Nowak and May~\cite{NoMa92} 
for the filled lattice in a dramatic way. 
The spatial chaos
is absent when the disorder is above a given amount, what is
reflected by the small number of active sites and the pinning
of the interfaces. It must be noticed that even without
disorder, we are not dealing
with a system presenting coarsening as the movement of the
interfaces is not ruled by surface tension as, e.g. in the
Ising model.

Besides presenting a dynamical transition in the presence of
disorder, dilution can even enhance the fraction of
cooperators in the population. Indeed, in the region
$4/3<b<3/2$, the relative density of cooperators in
the population can reach a value almost $40\%$ higher
than in the undiluted case, the maximum happening
at the transition point.

Some questions, however, are still open. For example, if the condition
of strong population viscosity is relaxed, 
disorder is no longer quenched and diffusion is allowed, what
happens with the cooperator groups? Obviously, the results
depend on the chosen rule for the diffusing elements, and
there are several biologically motivated rules. Work
is in progress in this direction. Moreover,
from the point of view of the study of persistence, it would
be interesting to study the behavior of the $P(t,t_w)$ above
$\rho^*$, for $1\ll t_w \ll t$ (instead of $t_w=0$) as well
as to compare the results presented here with the {\em site}
diluted Ising model.



\noindent
{\bf Acknowledgments:} We are grateful to H. Chat\'e, N. Lemke and
L. Peliti for discussions and suggestions and Y. Levin
for a critical reading of the manuscript. Work
partially supported by CNPq and PROPESQ-UFRGS.

\bibliographystyle{unsrt}

\end{multicols}
\end{document}